\documentclass{iopart}
\usepackage[T1]{fontenc}

 \expandafter\let\csname equation*\endcsname\relax
 \expandafter\let\csname endequation*\endcsname\relax

\usepackage{amsmath}
\usepackage{amssymb}
\usepackage{graphicx}
\usepackage{esint,url}



\def \tr {\text{Tr}}
\def \beq {\begin{equation}}
\def \edq {\end{equation}}
\def \bes {\begin{subequations}}
\def \eds {\end{subequations}}
\def \beqn {\begin{equation*}}
\def \edqn {\end{equation*}}

\def \dag {\dagger}

\def \veps {\varepsilon}

\def \calg {{\cal{G}}}
\def \calh {{\cal{H}}}

\def \calw {{\cal{W}}}
\def \cald {{\cal{D}}}
\def \hc {\text{H.c.}}
\def \bi {\bar{i}}

\begin{document}
\title{Engineering drag currents in Coulomb coupled quantum dots}
\author{Jong Soo Lim$^{1}$, David S\'anchez$^{2}$ and Rosa L\'opez$^{2}$}
\address{$^1$School of Physics, Korea Institute for Advanced Study, Seoul 130-722, Korea \\
$^2$Institute for Cross-Disciplinary Physics and Complex Systems IFISC (UIB-CSIC), E-07122 Palma de Mallorca, Spain}
\ead{david.sanchez@uib.es}

\begin{abstract}
The Coulomb drag phenomenon in a Coulomb-coupled double quantum dot system is revisited with a simple model
that highlights the importance of simultaneous tunneling of electrons.
Previously, cotunneling effects on the drag current in mesoscopic setups have been reported both theoretically and experimentally. 
However, in both cases the sequential tunneling contribution to the drag current was always present unless the drag level position were too far away from resonance.
Here, we consider the case of very large Coulomb interaction between the dots, whereby the drag current needs to be assisted by cotunneling events. As a consequence, a quantum coherent drag effect takes place.
Further, we demonstrate that by properly engineering the tunneling probabilities using band tailoring it is possible to control the sign of the drag and drive currents, allowing them to flow in parallel or antiparallel directions. We also show that the drag current can be manipulated by varying the drag gate potential and is thus governed by electron- or hole-like transport.
\end{abstract}

\maketitle

\section{Introduction}

In a system made of two nearby (electrically) isolated conductors where particles are prevented from tunneling into each other, 
a bias drop through one conductor can drag a current through the other conductor due to Coulomb interaction between them \cite{Rojo99, Narozhny15}.
The Coulomb drag effect was first suggested by Pogrebinskii \cite{Pogrebinskii77} and Price \cite{Price83} theoretically and
explained by energy and momentum transfer from the drive conductor to the drag conductor due to Coulomb mutual friction.
This phenomenon has become a quite useful toolbox
to probe electron-electron or electron-hole scattering mechanisms in many-body systems. 
At the early stage, the effect has been observed in 
semiconductor 2D-3D electron gas layers \cite{Solomon89}, 2D-2D electron gas layers \cite{Gramila91}, 2D electron gas-2D hole gas layers \cite{Sivan92} and then investigated under the influence of a perpendicular magnetic field \cite{Hill96, Rubel97, Lilly98, Feng98, Lok01, Kellogg02, Kellogg03, Muraki04, Tutuc09, Nandi12}. 
Since the Coulomb interaction is stronger in 1D than 2D, Coulomb drag between 1D-1D quantum nanowires has been also extensively explored \cite{Debray00, Debray01, Yamamoto02, Laroche11}.
More recently, interest has shifted to more exotic systems such as
graphene double layers \cite{Kim11, Gorbachev12, Kim12}, graphene double ribbons \cite{Chen13}, graphene-2D electron gases \cite{Gamucci14}, or double bilayer graphene heterostructures \cite{Lee16}.
In addition, Coulomb drag between
electrostatically coupled arrays of metalic tunnel junctions \cite{ave91,mat97}, between 
quantum wire and quantum dot \cite{Shimizu05}, between quantum dots \cite{Shinkai09}, or between quantum point contacts \cite{Khrapai07} has been demonstrated. In such systems translational symmetry is broken. Hence, Coulomb drag effect assisted by momentum transfer is no longer possible and only energy transfer between the two subsystems takes place. Energy exchange between the drive and drag subsystems leads to rectification of nonequilibrium fluctuations~\cite{lev08} in close analogy to the ratchet effect~\cite{astumian,mol10} with possible energy harvesting applications~\cite{hussein}.

Reducing the dimensionality of nanoscale devices has a two-fold advantage, namely: (i) the Coulomb interaction enhances by diminishing the system size, which leads to a major dragging effect; (ii) a very small drive voltage yields a drag current generated by the excess of noise of the drive system. Both properties may play significant roles in quantum information processing. Enhanced Coulomb interaction can entangle particles in a controlled way while drag currents can be used for quantum measurement purposes, where the drag subsystem serves as the detector. In this context, drag currents can be regarded as the back-action resulting from quantum measurements in the drive system. Importantly, these systems exhibit quantum coherence. Here, we demonstrate that quantum coherence is crucial in the strongly interacting mesoscopic Coulomb drag effect.  

Reference~\cite{Sanchez10} proposes a simple mesoscopic setup that consists of two parallel Coulomb-coupled quantum dots attached to four separate normal electron reservoirs to generate the Coulomb drag current. Using this device, it is possible to test fundamental fluctuation relations among nonlinear transport coefficients. 
Both quantum dots are closely spaced so that a strong interdot Coulomb interaction is expected and the setup is thus a natural platform to observe Coulomb drag. Interdot tunneling is forbidden. The setup is analyzed employing the master equation approach, considering only sequential tunneling rates.
Dot charge-state dependent tunnel barriers are key ingredients in the prediction of unidirectional drag currents, independently of the direction of the drive bias drop.
A recent experiment \cite{Keller16} realizes the setup using a lithographically patterned AlGaAs/GaAs heterostructure and detects  unidirectional drag currents as anticipated. However and in contrast to the theory of Ref.~\cite{Sanchez10},
the experiment shows at very low temperatures the importance of including cotunneling processes,
as demonstrated with a theoretical model that shows excellent agreement with the experimental data \cite{Keller16}.
The role of cotunneling processes is also emphasized in Ref.~\cite{Kaasbjerg16} where a similar system is considered but with graphene reservoirs instead of normal electron reservoirs~\cite{bis15}. In both cases, a four-charge state model within a master equation description serves as a theoretical basis to describe the Coulomb drag effect. It is proven that nonlocal cotunneling processes contribute to the drag current to the same order as sequential tunneling events, the former being the main contribution when the dot level for the drag subsystem lies well far away from the resonant condition. In such situation, the drag current is mainly driven by high-order tunneling processes~\cite{Keller16,Kaasbjerg16}. Since both sequential and cotunneling events contribute to the drag effect in the formulations of Refs.~\cite{Keller16,Kaasbjerg16}, it would be highly desirable to put forward a theoretical model that leads to cotunneling driven drag currents as opposed to the purely sequential regime discussed earlier~\cite{Sanchez10}.

\begin{figure}[t]
\centering
\includegraphics[width=0.65\textwidth,angle=0,clip]{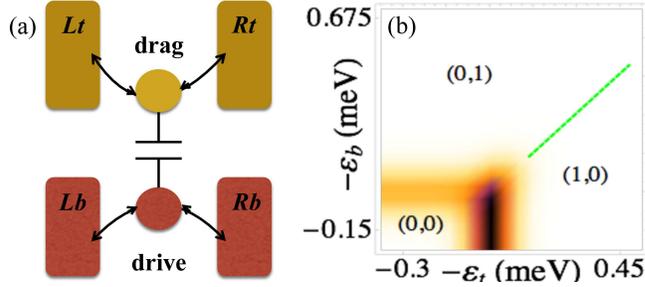}
\caption{(a) Sketch of our device. Top dot acts as the drag subsystem held at 
equilibrium whereas an electrical current is driven across the bottom dot (the drive subsystem).
Both dots are tunnel connected to reservoirs  as indicated. 
Coulomb interaction between the two dots (here, represented with a capacitor)
is necessary for the drag effect to happen. (b) Stability diagram as a function of the dot top level ($\veps_t$) and 
the bottom dot level ($\veps_b$). Dot occupations are denoted with $(n_t,n_b)$. 
The borders between regions with constant occupations are given by conductance values.
In the limit of strong interdot interactions, there exists a unique triple point where charge fluctuates among the states $(0,0)\leftrightarrows(0,1)\leftrightarrows(1,0)$. The dotted line is for eye-guiding purposes.}
\label{fig:stability}
\end{figure}

Our aim here is then to analyze a simple model that totally suppresses drag currents induced by sequential tunneling alone.
This is achieved by considering a mutual Coulomb interaction sufficiently large to discard the doubly occupied charge state.
Indeed, a three-state model with solely sequential tunneling processes is unable to create drag currents. This finding emphasizes the importance of cotunneling in the physics of the drag effect.
Below, we demonstrate that three states represent the minimum configuration that generates drag current around the triple point of interacting dot setups. We also show that the drag transport in this configuration must be assisted by cotunneling mechanisms unlike the four-state model employed in previous works~\cite{Sanchez10,Keller16,Kaasbjerg16}, in which sequential contributions to the drag current are always present.
The four-state model can be also employed to cancel the purely sequential drag currents but only if temperature and voltage are very low~\cite{Kaasbjerg16}.
In contrast, the three-state model proposed in this work is much easier to handle and exactly nullifies the pure sequential part.
Based on this simple model, we prove that the behavior and sign of the mesoscopic drag current can be controlled externally by properly engineering the energy dependence of the tunneling barriers. This can be understood in terms of an intuitive picture of electron- and hole-type transport.

The system under consideration is plotted in Fig.~\ref{fig:stability}(a).
Two parallel coupled quantum dots are attached to four normal electronic reservoirs.
Particle flow is indicated with arrows while electron-electron
interaction occurs between the dots only (the reservoirs are assumed to be massive electrodes with good screening properties).
The stability diagram  for this setup is shown in  Fig.~\ref{fig:stability}(b).  The diagram is obtained by summing the linear conductances $G_i = (dI_i/dV_i)_{V_i=0}$ ($i=t,b$) through the drag (top, $t$) and drive (bottom, $b$) dots and plotted as a function of the top ($\veps_t$) and the bottom ($\veps_b$) gate voltages. Here, $I_i$ ($V_i$) is the current (voltage) across the subsystem $i=t,b$ and $(n_t,n_b)$
denotes the dot charge configurations, where $n_t$ and $n_b$ are the number of electrons on dot $t$ and $b$.
When the Coulomb interaction strength is very large the resulting stability diagram shown in Fig.~\ref{fig:stability}(b) displays only one triple point where the doubly occupied configuration $(1,1)$ never takes place, $(1,0)$ and $(0,1)$ being equally probable.
We are interested in the triple point because experimental observations~\cite{Keller16,bis15} find the largest drag currents in that region of the stability diagram.

\section{Theory}

The model Hamiltonian
\begin{equation}
\calh = \calh_D + \calh_C + \calh_T
\end{equation}
has three contributions. First,
\begin{equation}
\calh_D = \sum_{i=t/b} \veps_i d_i^{\dag}d_i + Ud_t^{\dag}d_td_b^{\dag}d_b
\end{equation}
describes the quantum dots.
The operator $d_i^{\dag}$ ($d_i$) creates (annihilates) an electron with energy $\veps_i$ on the dot $i$.
The interdot charging energy is denoted with $U$.
By adjusting gate voltages appropriately, the direct tunneling between quantum dots is forbidden
and can be safely neglected~\cite{Keller16}.
Second,
\begin{equation}
\calh_C =  \sum_{\alpha=L/R,i,k} \veps_{\alpha ik} c_{\alpha ik}^{\dag}c_{\alpha ik}
\end{equation}
depicts the reservoir Hamiltonian where $c_{\alpha ik}^{\dag}$ ($c_{\alpha ik}$) creates (annihilates) an electron 
with wavevector $k$ and energy $\veps_{\alpha ik}$ in the reservoir $\alpha i=\{Lt,Lb,Rt,Rb\}$ [Fig.~\ref{fig:stability}(a)].
Finally,
\begin{equation}
\calh_T = \sum_{\alpha,i,k}\left(t_{\alpha ik}c_{\alpha ik}^{\dag}d_i + \hc\right)
\end{equation}
characterizes the coupling between the top and bottom dots and the reservoirs via the tunneling amplitudes $t_{\alpha ik}$.

We focus on the dot states, tracing out the reservoir degrees of freedom.
In our system and for $U\to\infty$, there are three possible dot states $|n\rangle = \{|0\rangle, |t\rangle, |b\rangle\}$.
The state $|0\rangle \equiv (0,0)$ implies that both dots are empty, while $|t\rangle \equiv (1,0)$ ($|b\rangle \equiv (0,1)$) indicates that the top (bottom) dot is occupied.
To lowest order in $\calh_T$, the transition (sequential tunneling) rate from $|m\rangle$ to $|n\rangle$
entering or leaving the reservoir $\alpha i$
is given by \cite{Bruus04}
\begin{align} \label{eq_seq}
\calw_{mn}^{\alpha i} = \frac{2\pi}{\hbar}\tr_C\left[F_{m_C}|\langle n_C|\langle n|\calh_T|m\rangle|m_C\rangle|^2\right]
\delta(E_{m}+E_{m_C}-E_{n}-E_{n_C})\,,
\end{align}
where $\tr_C$ designates a trace over the reservoir degrees of freedom, $F_{m_C}$ is the thermal distribution function for the reservoir state $m_C$ with energy $E_{m_C}$,
and $E_m$ and $E_n$ are the energies of the states $|m\rangle$ and $|n\rangle$.
The second-order rate in $\calh_T$ describes the cotunneling processes from $|m\rangle$ to $|n\rangle$ and is calculated as
\begin{align} \label{eq_cot}
\gamma_{mn}^{\alpha i\beta j} = \frac{2\pi}{\hbar}\tr_C\left[F_{m_C}|\langle n_C|\langle n|\calh_T\calg_0\calh_T|m\rangle|m_C\rangle|^2\right] 
\delta(E_{m}+E_{m_C}-E_{n}-E_{n_C})\,,
\end{align}
where  $\calg_0$ is the resolvent operator given by $\calg_0 = \frac{1}{E_m - \calh_0}$ with $\calh_0 = \calh_D + \calh_C$.
Conceptually, one might go further by including higher-order tunneling processes in $\calh_T$
but this is out of the scope of the present work. Our goal is to consider quantum coherent processes to leading order.
This approximation is valid provided that the background temperature is not very low ($k_B T > \Gamma_0$, where
$\Gamma_0$ is the tunnel barrier strength energy to be specified below).

Let $P_n$ be the probability for the state $|n\rangle$. Within the master equation approach, we can write
\bes
\begin{align}
\frac{d}{dt} P_0 &= -\left(\calw_{0t} + \calw_{0b}\right)P_0 + \calw_{t0}P_t + \calw_{b0}P_b \,,
\\
\frac{d}{dt} P_t &= \calw_{0t}P_0 -\left(\calw_{t0} + \gamma_{tb}\right)P_t + \gamma_{bt}P_b  \,,
\\
\frac{d}{dt} P_b &= \calw_{0b}P_0 + \gamma_{tb}P_t - \left(\calw_{b0} + \gamma_{bt}\right)P_b \,,
\end{align}
\label{eq:fpopupd}
\eds
where $\calw_{mn} = \sum_{\alpha i} \calw_{mn}^{\alpha i}$
and $\gamma_{mn} = \sum_{\alpha i,\beta j} \gamma_{mn}^{\alpha i\beta j}$.
Importantly, the singly occupied states $|t\rangle$ and $|b\rangle$ are interconnected thanks to the cotunneling rates.
In contrast, sequential tunneling processes always involve the empty state $|0\rangle$.
We stress that the model described by Eq.~\eqref{eq:fpopupd} can be obtained from that of Ref.~\cite{Keller16}
first by carefully removing the doubly occupied state and only then taking the limit $U\to\infty$
in the tunneling rate expressions.

The stationary probabilities (${\rm st}$) can be obtained from Eq.~\eqref{eq:fpopupd} in the limit $t\to\infty$ with
the aid of the probability conservation law $\sum_n P_n = 1$. The full expressions for the rates can be found in~\ref{app:Rates}. 
Let us focus on the current flowing into the left top $(Lt)$ reservoir ($I_{\rm drag} \equiv I_{Lt}$),
\begin{equation}
I_{\rm drag} = -e\left[\calw_{0t}^{Lt}P_0^{\rm st}
- \left(\calw_{t0}^{Lt} + \gamma_{tb}^{LbLt} + \gamma_{tb}^{RbLt}\right)P_t^{\rm st}
\right. \left. + \left(\gamma_{bt}^{LtLb} + \gamma_{bt}^{LtRb}\right)P_b^{\rm st}
\right] \,,
\label{eq:FUIdrag}
\end{equation}
where $e$ denotes the elementary positive electric charge.
Crucially, when cotunneling processes are neglected the drag current identically vanishes 
and the doubly occupied state must be then considered.
In the case of infinite charging energy treated here, the doubly occupied state
is energetically forbidden and then cotunneling is required to create drag currents.

On the other hand, the drive current ($I_{drive} \equiv I_{Lb}$) is
\begin{multline}
\frac{I_{\rm drive}}{-e} = (\calw_{0b}^{Lb} - \gamma_{00}^{RbLb} + \gamma_{00}^{LbRb})P_0^{\rm st}
+ (\gamma_{tb}^{LbLt}
+ \gamma_{tb}^{LbRt})P_t^{\rm st}\\
- (\calw_{b0}^{Lb} + \gamma_{bb}^{RbLb} - \gamma_{bb}^{LbRb} + \gamma_{bt}^{LtLb} + \gamma_{bt}^{RtLb})P_b^{\rm st}\,.\label{eq:FUIdrive}
\end{multline}
Unlike Eq.~\eqref{eq:FUIdrag}, Eq.~\eqref{eq:FUIdrive} can be nonzero if the $\gamma$'s are nullified since
a voltage is present in the drive subsystem.

\begin{figure}[t!]
\centering
\includegraphics[width=0.75\textwidth,angle=0,clip]{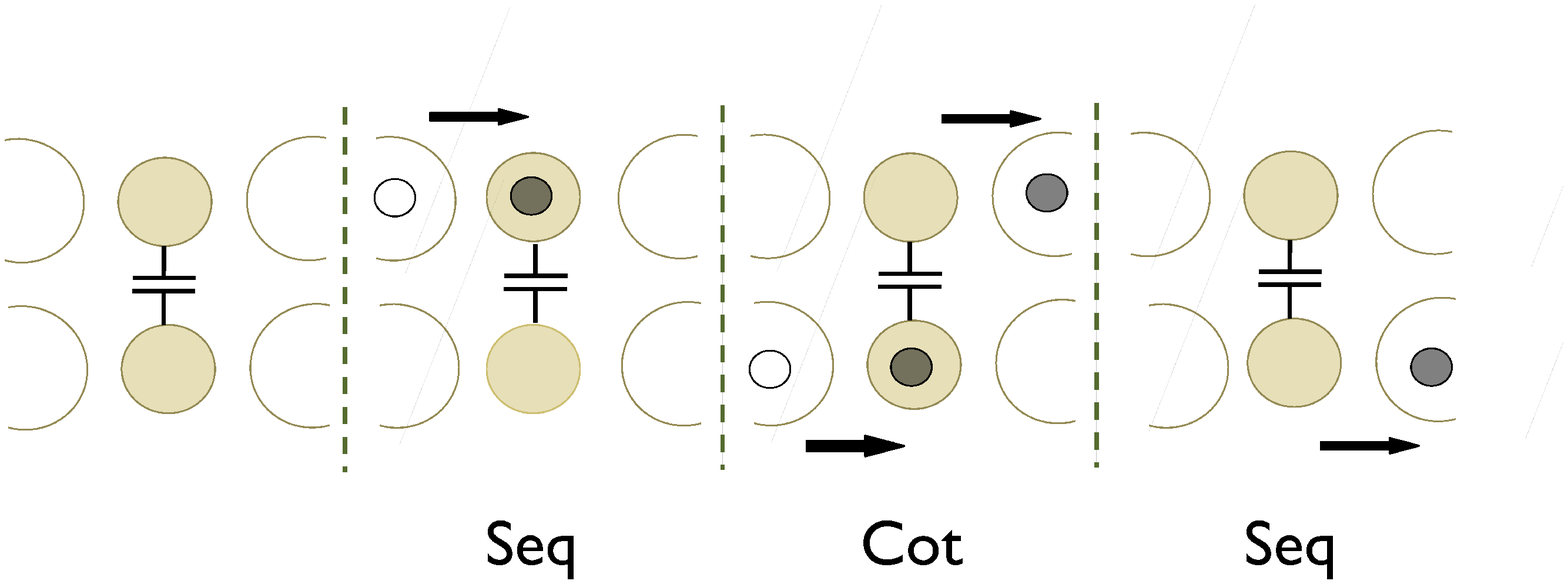}
\caption{Characteristic transport sequence involving the drag of an electron through the top dot. Empty (filled)
circles represent electrons leaving (arriving at) the different parts of the system (reservoirs or dots).
Arrows indicate the transport direction. We consider a positive bias applied to the drive subsystem. Hence,
we only indicate right-going processes for the lower dot.
While the second and four panels depict sequential processes (one arrow, i.e., first order in tunneling),
the third panel illustrates a cotunneling event (two arrows, i.e., second order in tunneling). The latter
process is essential to drive a drag current within our model.}
\label{fig:sequence}
\end{figure}

Calculating the stationary probabilities explicitly (see~\ref{app:Currents}), we find
\begin{align}\label{eq_Idrag}
I_{\rm drag}^{\rm (seq)} 
\propto \Gamma_{Lt}(\veps_t)\Gamma_{Rt}(\veps_t) \left[f_{Rt}(\veps_t) - f_{Lt}(\veps_t)\right] 
\sum_{\alpha} \Gamma_{\alpha b}(\veps_b) \left[1 - f_{\alpha b}(\veps_b)\right]\,,
\end{align}
where $f_{\alpha i}(\veps)=1/\left[1+e^{(\veps-\mu_{\alpha i})/k_BT}\right]$ is the Fermi function
with $\mu_{\alpha i}$
the electrochemical potential of lead ${\alpha i}$. The dot electrochemical potentials are 
to be evaluated using an electrostatic model discussed in detail in~\ref{app:EM}.
This model is especially appealing for comparison with the experiment~\cite{Keller16}
and renders our theory gauge invariant.
The hybridization widths $\Gamma_{\alpha i}(\veps) = 2\pi\sum_k t_{\alpha ik}^2\delta(\veps_{\alpha ik} - \veps) = 2\pi t_{\alpha i}^2(\veps)\rho_{\alpha i}(\veps)$ depend on the tunnel amplitude $t_{\alpha i}$ and density of states (DOS) $\rho_{\alpha i}$.
Since $\mu_{Lt} = \mu_{Rt}$ in the drag system, we have $f_{Lt}(\veps) = f_{Rt}(\veps) \equiv f_t(\veps)$ leading to a zero drag current
($I_{\rm drag}^{\rm (seq)} =0$) in the sequential tunneling regime
within our three-state model (which corresponds physically to the case $U\to \infty$),
even if the tunneling barriers are asymmetric and energy dependent.
As a consequence, cotunneling events are strictly needed to generate drag currents in the configuration of three charge states.

Consider the sequence $|0\rangle \to |t\rangle \to |b\rangle \to |0\rangle $ illustrated in Fig.~\ref{fig:sequence}, where an electron is transported from left to right in the drag subsystem.
The corresponding probability is proportional to $ \calw_{0t}^{Lt} \gamma_{tb}^{\alpha bRt} \calw_{b0}^{\beta b}$.
The probability for the reversed process reads $\calw_{0t}^{Rt} \gamma_{tb}^{\alpha bLt} \calw_{b0}^{\beta b}$.
As a consequence, the net current,
\begin{equation}
I_{\rm drag} \propto \sum_{\alpha,\beta} \left(\calw_{0t}^{Lt} \gamma_{tb}^{\alpha bRt} \calw_{b0}^{\beta b} - \calw_{0t}^{Rt} \gamma_{tb}^{\alpha bLt} \calw_{b0}^{\beta b}\right)
\end{equation}
is clearly assisted by quantum coherent processes.
We observe in Fig.~\ref{fig:sequence} that two sequential transitions
[i.e., $|0\rangle \to |t\rangle$ (second panel)
and $b\rangle \to |0\rangle$ (fourth panel)] are still present but they are not sufficient
to drive a drag current since the middle transition ($t\rangle \to |b\rangle$, third panel)
arises solely from cotunneling.

In terms of the explicit expressions for the tunneling rates \cite{Averin94,Turek02},
the drag current takes the form
\begin{align}\label{I_dragcot}
I_{\rm drag} &\propto \sum_{\alpha,\beta} \int d\veps~\left|\frac{1}{\veps-\veps_b}\right|^2
[\Gamma_{Lt}(\veps_t)\Gamma_{Rt}(\veps+\veps_t-\veps_b) - \Gamma_{Rt}(\veps_t)\Gamma_{Lt}(\veps+\veps_t-\veps_b)]\nonumber \\
&\times\Gamma_{\alpha b}(\veps)\Gamma_{\beta b}(\veps_b)
 f_t(\veps_t)f_{\alpha b}(\veps)\left[1-f_t(\veps+\veps_t-\veps_b)\right]\left[1-f_{\beta b}(\veps_b)\right] \,.
\end{align}

We remark that $I_{\rm drag}$ in Eq.~\eqref{I_dragcot} vanishes when $\Gamma_{\alpha i}(\veps) = \Gamma_{\alpha i}$
and thus energy-dependent tunnel barriers are necessary~\cite{Sanchez10,Keller16,Kaasbjerg16}.
We envisage that one of the tunnel barriers (top left, $Lt$) has a Lorentzian profile,
\beq
\Gamma_{Lt}(\veps) = \frac{\Gamma_{Lt}}{1 + \left[(\veps-\mu_{Lt})/D\right]^2}\,,
\label{eq:Lorentzian}
\edq
and the other tunnel barriers are constant, i.e., $\Gamma_{Rt}(\veps) = \Gamma_{Rt}$, $\Gamma_{Lb}(\veps) = \Gamma_{Lb}$, $\Gamma_{Rb}(\veps) = \Gamma_{Rb}$. 
This model can be realized with constant tunnel amplitudes ($t_{\alpha i}(\veps) = t_{\alpha i}$)
while the DOS for the left top reservoir $\rho_{Lt}(\veps)$ is Lorentzian and the other reservoirs have flat bands.
Tailoring of the band structure in the leads can be achieved with different materials~\cite{bandstructure}.
In the following we present our results for the drive and drag currents and discuss the tunability of the drag current depending
on the system parameters. 

\begin{figure}[t]
\centering
\includegraphics[width=0.6\textwidth]{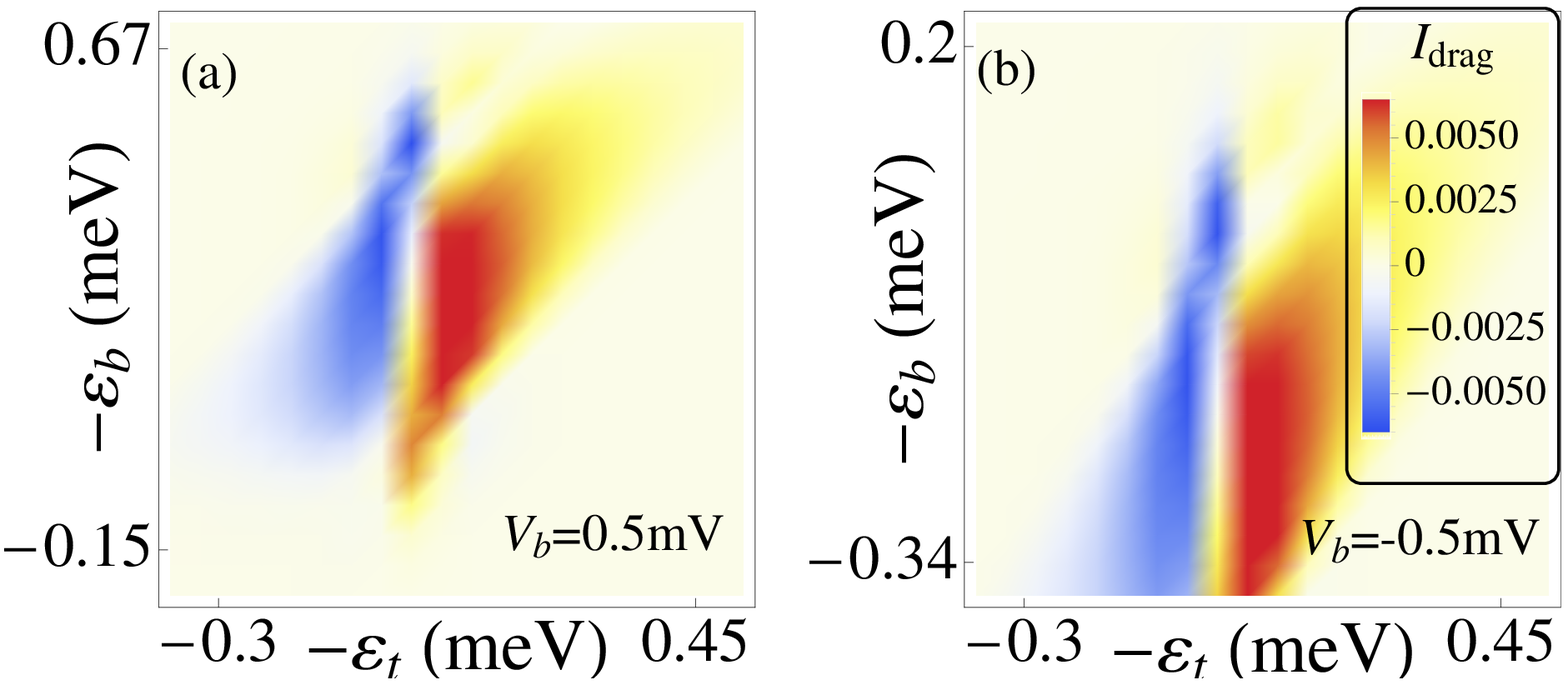}
\includegraphics[width=0.6\textwidth]{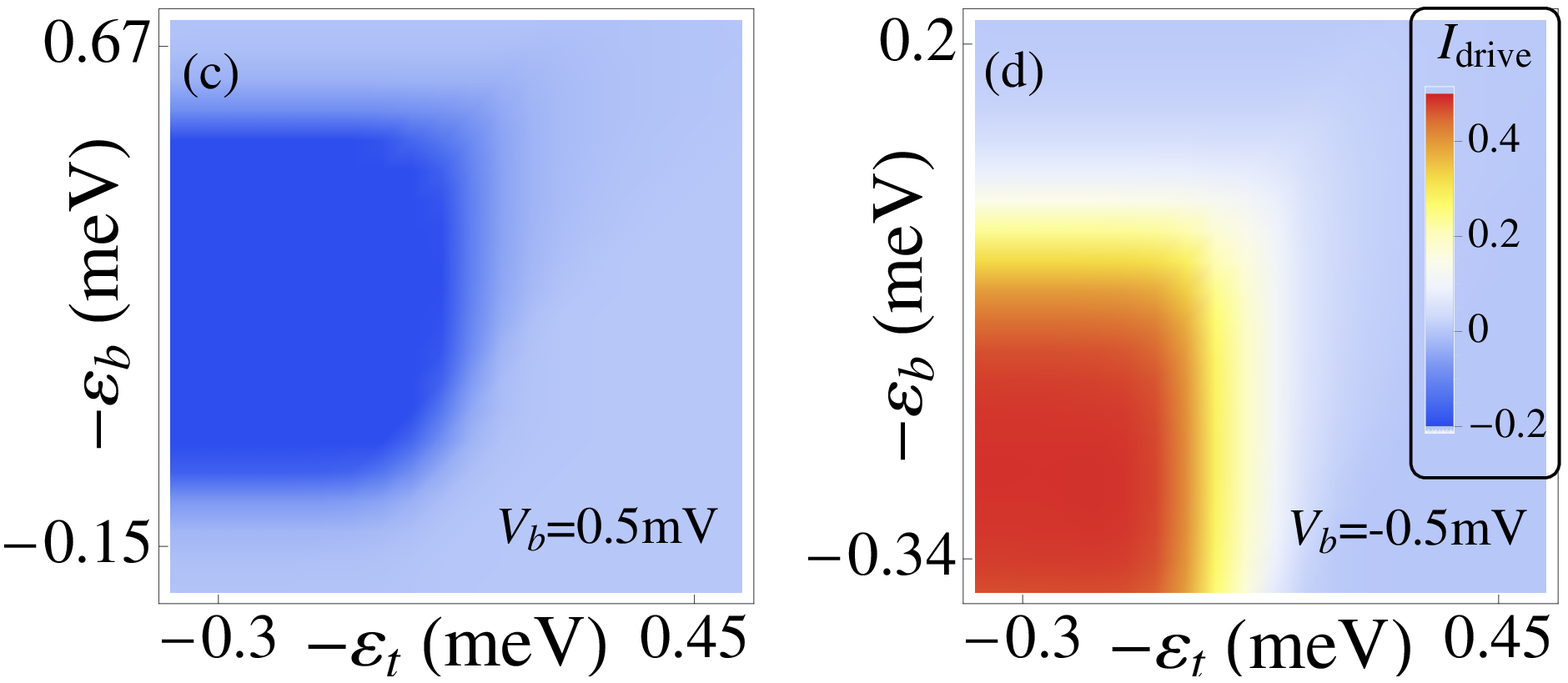}
\caption{Drag current $I_{\rm drag}$ (top panels) and drive current $I_{\rm drive}$ (bottom panels) as a function of top ($\veps_t$) and bottom ($\veps_b$) dot level positions. The applied drive bias voltage is $V_b =0.5$~mV (a,c) and $V_b = - 0.5$~mV (b,d). Current is expressed in units of $e\Gamma_0/\hbar$. For reference purposes, we find a maximum value for the ratio $|I_{\rm drag}/I_{\rm drive}|$ of the order of $10^{-2}$.
}
\label{fig:dragdrive}
\end{figure}

\section{Results}\label{sec_results}

\begin{figure}[t]
\centering
\includegraphics[width=0.14\textwidth,angle=-90,clip]{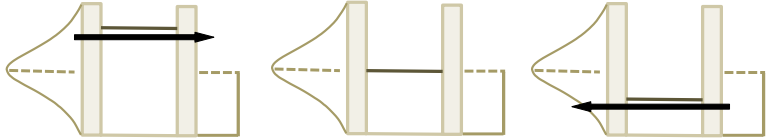}
\caption{Sketch of the drag dot for $\mu_{Lt}=\mu_{Rt}$ (dashed lines) and three different level positions
(filled dark lines).
The left (right) reservoir displays a Lorentzian (flat) DOS, which determines the current direction
(long arrows).}
\label{fig:sign}
\end{figure}

Throughout this paper, we take $\Gamma_{Lt} = \Gamma_{Rt} = (47/15)\Gamma_0$ and $\Gamma_{Lb} = \Gamma_{Rb} = \Gamma_0$ where $\Gamma_0=7.5$~$\mu$eV is the unit of energy.
These values were extracted from the experiment \cite{Keller16}.
The temperature is set to $k_BT = 5\Gamma_0$ and $D=10\Gamma_0$.
We evaluate the drag ($I_{\rm drag}$) and drive ($I_{\rm drive}$) currents as a function of $\veps_t$ and $\veps_b$
for a drive voltage $V_b \ne 0$.
In Figs.~\ref{fig:dragdrive}(a) and (b), we observe that $I_{\rm drag}$
is strongest along the (green) dotted line shown in Fig.~\ref{fig:stability}(b).
Therefore, drag currents are intimately related to charge fluctuations since
$I_{\rm drag}\neq 0$ for the sequence depicted in Fig.~\ref{fig:sequence}, in which case
the three states $|0\rangle$, $|t\rangle$, $|b\rangle$ have similar probabilities
and as a consequence $I_{\rm drag}$ is maximal around the triple point in Fig.~\ref{fig:stability}(b).
Reversing the direction of $V_b$ does not change the direction of $I_{\rm drag}$
but merely shifts the drag regions down following the bias window~\cite{Keller16}. Surprisingly enough,
unlike the experimental results \cite{Keller16} our drag current changes sign when $\veps_t$ is tuned. 
The phenomenon is illustrated in Fig.~\ref{fig:sign}. When $\veps_t>\mu_L$ (left panel) thermally excited electrons
from the left reservoir can tunnel to the right lead and a negative current is then obtained
(our flux sign convention states that the current is negative when it leaves the reservoir).
If $\veps_t<\mu_L$ (right panel) the $I_{\rm drag}$ direction is reversed because electrons
below the Fermi energy in the right lead tend to fill the available states in the left reservoir,
thus creating hole-like transport. Obviously, for $\veps_t=\mu_L$ (middle panel) electron- and hole-like
fluxes compensate each other and the net current is zero [see the border line between the blue and red regions
in Fig.~\ref{fig:dragdrive}(a,b)]. This resembles the thermoelectric effect in quantum dots but the difference here is that here the thermal bias is zero. The common origin for both effects is the breaking of electron-hole symmetry.

The discussed effect implies that we can engineer $I_{\rm drag}$ by gate tuning $\veps_t$
with a suitable model for the electronic DOS at the reservoirs. 
This agrees with the findings of Ref.~\cite{Kaasbjerg16}. The difference is that Ref.~\cite{Kaasbjerg16} apply gate voltages
to graphene reservoirs while in our double quantum dot setup we tune the energy level of the dots.
Incidentally, the region in which there is an appreciable drag current is extended to $\veps_t$ values far away from resonance, as in the experiment~\cite{Keller16}. This is due only to cotunneling processes since a purely sequential model yields sharp boundaries~\cite{Sanchez10}.

In Figs.~\ref{fig:dragdrive}(c) and (d) we plot the drive current as a function of the level position for a nonzero $V_b$.
For $V_b>0$ ($V_b<0$) the current is negative (positive), as expected.
In both cases the current is nonzero only when the $\veps_b$ is within the bias window
[i.e., when $-eV_b-\veps_b^0\lesssim \veps_b\lesssim \veps_b^0$ where $\veps_b^0\simeq 0$~meV signals the transition $(0,0)\to(0,1)$ in Fig.~\ref{fig:stability}(b)].
Interestingly, $I_{\rm drive}$ also depends on $\veps_t$. This can be understood as follows.
When the top dot level falls below the Fermi energy, the drag dot becomes occupied and
transport across the drive dot remains blocked due to the strong electron-electron interaction
between the dots (gating effect). Importantly, for a wide range of $\veps_t$ both currents (drag and drive)
run in parallel if $V_b>0$ [cf. Figs.~\ref{fig:dragdrive}(a) and (c)] but flow in antiparallel
directions if $V_b<0$ [cf. Figs.~\ref{fig:dragdrive}(b) and (d)].
This results in an additional degree of tunability by controlling $\veps_d$ and $V_b$ independently. 

\section{Summary}

We have considered a three-state model to propose a device for the observation of a drag effect that needs the presence of quantum coherent events. 
This is the minimal description that generate drag currents around the triple point of Coulomb coupled double quantum dots. A clever design for the attached reservoir DOS leads to a powerful tunability for the drive and drag currents. Thus, we have illustrated our findings with a model in which the left top coupling has a Lorentzian density profile and the other couplings are constant.
We have demonstrated that the drag current flows in parallel or in opposite direction to the drive current by reversing the sign of the bias voltage and that the drag changes changes sign when the dot energy level is varied with an external gate potential. Our proposal hence opens a new route for the creation and manipulation of coherent drag currents in nanoscale conductors. 

\section*{Acknowledgments}

This work was supported by MINECO under grant No.\ FIS2014-52564.

\appendix

\section{Calculation of the transition rates} \label{app:Rates}

\subsection{Sequential tunneling}

In Eq.~\eqref{eq_seq}
$F_{m_C}$ is given by
\beq
F_{m_C} = \frac{e^{-\beta\calh_C}}{\tr_C \left[e^{-\beta\calh_C}\right]} \,.
\edq
As an example, we evaluate the rate $\calw_{0i}^{\alpha i}$ from $|0\rangle$ to $|i\rangle$.
Since the final state is
\beq
|n\rangle|n_C\rangle = d_i^{\dag}c_{\alpha ik} |0\rangle|m_C\rangle \,,
\edq
the rate can be written as
\begin{align}
\calw_{0i}^{\alpha i} = \frac{2\pi}{\hbar} \sum_k\sum_{m_C} F_{m_C}\left|\langle m_C|\langle 0|c_{\alpha ik}^{\dag}d_id_i^{\dag}c_{\alpha ik}|0\rangle|m_C\rangle\right|^2 
 \delta(\veps_{\alpha ik}-\veps_i) \,.
\end{align}

We note that
\beq
\sum_{m_C} F_{m_C} |\langle m_C|c_{\alpha ik}^{\dag}c_{\alpha ik}|m_C\rangle|^2
= f_{\alpha i}(\veps_{\alpha ik}) \,.
\label{eq:fermi}
\edq
Equation~\eqref{eq:fermi} follows from the fact that $c_{\alpha ik}^{\dag}c_{\alpha ik}$ can only take the values $0$ or $1$.
Therefore, we obtain 
\beq
\calw_{0i}^{\alpha i} = \frac{1}{\hbar} \Gamma_{\alpha i}(\veps_i)f_{\alpha i}(\veps_i) \,,
\label{eq:seq0i}
\edq
where
\beq
\Gamma_{\alpha i}(\veps_i) = 2\pi\sum_k t_{\alpha ik}^2 \delta(\veps_{\alpha ik}-\veps_i) \,.
\edq
The sequential tunneling rate from $|i\rangle$ to $|0\rangle$ can be similarly found,
\beq
\calw_{i0}^{\alpha i} = \frac{1}{\hbar} \Gamma_{\alpha i}(\veps_i)\left[1-f_{\alpha i}(\veps_i)\right] \,.
\label{eq:seqi0}
\edq

\subsection{Cotunneling}

The cotunneling rate from $|m\rangle$ to $|n\rangle$ is given by Eq.~\eqref{eq_cot}.
For energy dependent tunnel broadenings, we find
\bes
\begin{align}
\gamma_{00}^{\alpha i\beta i} 
&= \frac{1}{2\pi\hbar} \int d\veps~\left|\frac{1}{\veps-\veps_i+i\eta}\right|^2 \Gamma_{\alpha i}(\veps)\Gamma_{\beta i}(\veps) f_{\alpha i}(\veps)\left[1 - f_{\beta i}(\veps)\right] \,,
\\
\gamma_{ii}^{\alpha i\beta i} 
&= \frac{1}{2\pi\hbar} \int d\veps~\left|\frac{1}{\veps-\veps_i+i\eta}\right|^2 \Gamma_{\alpha i}(\veps)\Gamma_{\beta i}(\veps)
f_{\alpha i}(\veps)\left[1 - f_{\beta i}(\veps)\right] \,,
\\
\gamma_{i\bi}^{\alpha\bi\beta i} 
&= \frac{1}{2\pi\hbar} \int d\veps~\left|\frac{1}{\veps-\veps_{\bi}+i\eta}\right|^2
\Gamma_{\alpha\bi}(\veps)\Gamma_{\beta i}(\veps+\veps_i-\veps_{\bi})
f_{\alpha\bi}(\veps)\left[1 - f_{\beta i}(\veps+\veps_i-\veps_{\bi})\right] \,,
\end{align}
\label{eq:CoTunL}
\eds
where $\bi = b/t$ for $i = t/b$.
Since the intermediate virtual states acquire a finite lifetime during the tunneling processes, we added a small positive parameter $\eta$.
We note that in contrast to Ref.~\cite{Keller16}
we here explicitly write down the expression for
$\gamma_{00}^{\alpha i\beta i}$ since it enters
the expression for $I_{\rm drive}$. In Ref.~\cite{Keller16}
the calculation of this rate is not needed because
only $I_{\rm drag}$ is shown.

We now use the identity
\beq
f_a(\veps_i)\left[1 - f_b(\veps_j)\right] 
= \frac{1}{2} n_B(\veps_i - \veps_j - \mu_a + \mu_b) \left[\tanh\left(\beta(\veps_i-\mu_a)/2\right) - \tanh\left(\beta(\veps_j-\mu_b)/2\right) \right] \,,
\edq
where $n_B(x) = 1/(\exp(\beta x)-1)$.
Here, $\tanh(z)$ can be expressed as 
\beq
\tanh(z) = -\frac{i}{\pi} \left[ \Psi\left(\frac{1}{2} + i\frac{z}{\pi}\right) - \Psi\left(\frac{1}{2} - i\frac{z}{\pi}\right) \right] \,,
\label{eq:tanhz}
\edq
where $\Psi(z)$ denotes the digamma function.
Employing the fact that $\Psi(\frac{1}{2}+i\frac{z}{2\pi})$ has the poles in the upper halfplane at the locations $z=i\pi(1/2+n)$, where $n\in N$, while the poles of $\Psi(\frac{1}{2}-i\frac{z}{\pi})$ are in the lower halfplane at the locations $z=-i\pi(1/2+n)$,
the integrals in Eqs.~\eqref{eq:CoTunL} can then be easily calculated.
In the final stage, we expand the results in powers of $\eta$. Those terms scaling as $1/\eta$ are removed since they are indeed sequential tunneling processes and cannot be counted twice. Then, we take the limit $\eta\to 0$ and find
\bes
\beq
\gamma_{00/bb}^{\alpha b\beta b} 
= \frac{\beta}{4\pi^2\hbar} \Gamma_{\alpha b}\Gamma_{\beta b} n_B(\mu_{\beta b}-\mu_{\alpha b})
\Im\left[\Psi^{(1)}\left(\frac{1}{2} + i\beta\frac{\veps_b-\mu_{\alpha b}}{2\pi}\right)
- \Psi^{(1)}\left(\frac{1}{2} + i\beta\frac{\veps_b-\mu_{\beta b}}{2\pi}\right)\right] \,,
\edq
\begin{multline}
\gamma_{tb}^{\alpha b Lt}
= \frac{\Gamma_{\alpha b}\Gamma_{Lt}}{2\pi\hbar} n_B(\veps_b - \veps_t + \mu_{Lt}-\mu_{\alpha b}) 
\\
\times\left\{\frac{D^2}{D^2 + (\veps_t-\mu_{Lt})^2} \Im \left[ \frac{2i(\veps_t-\mu_{Lt})}{D^2+(\veps_t-\mu_{Lt})^2} \left( \Psi\left(\frac{1}{2} + i\beta\frac{\veps_b-\mu_{\alpha b}}{2\pi}\right) -\Psi\left(\frac{1}{2} + i\beta\frac{\veps_t-\mu_{Lt}}{2\pi}\right)\right)
\right.\right.
\\
\left.
+ \frac{\beta}{2\pi} \left( \Psi^{(1)}\left(\frac{1}{2} + i\beta\frac{\veps_b-\mu_{\alpha b}}{2\pi}\right)
-\Psi^{(1)}\left(\frac{1}{2} + i\beta\frac{\veps_t-\mu_{Lt}}{2\pi}\right) \right) \right]
\\
\left.
+ \frac{1}{D} \Im \left[ \frac{D^2}{(\veps_u-\mu_{Lt}+iD)^2}
\left( \Psi\left(\frac{1}{2} + \frac{\beta D}{2\pi} + i\beta\frac{\veps_b-\veps_t-\mu_{\alpha b}+\mu_{Lt}}{2\pi}\right)
-\Psi\left(\frac{1}{2} + \frac{\beta D}{2\pi} \right) \right)\right] \right\} \,,
\end{multline}
\begin{multline}
\gamma_{bt}^{Lt\alpha b} = \frac{\Gamma_{Lt}\Gamma_{\alpha b}}{2\pi\hbar} n_B(\veps_t-\veps_b+\mu_{\alpha b}-\mu_{Lt})
\\
\times\left\{\frac{D^2}{D^2+(\veps_t-\mu_{Lt})^2} \Im \left[
\frac{2i(\veps_t-\mu_{Lt})}{D^2+(\veps_t-\mu_{Lt})^2} \left(\Psi\left(\frac{1}{2} + i\beta\frac{\veps_t-\mu_{Lt}}{2\pi}\right) - \Psi\left(\frac{1}{2} + i\beta\frac{\veps_b-\mu_{\alpha b}}{2\pi}\right) \right) \right.\right.
\\
\left. +
\frac{\beta}{2\pi} \left(\Psi^{(1)}\left(\frac{1}{2} + i\beta\frac{\veps_t-\mu_{Lt}}{2\pi}\right) - \Psi^{(1)}\left(\frac{1}{2} + i\beta\frac{\veps_d-\mu_{\alpha b}}{2\pi}\right) \right) \right]
\\
\left. +
\frac{1}{D} \Im \left[\frac{D^2}{(\veps_u-\mu_{Lt}+iD)^2} 
\left(\Psi\left(\frac{1}{2} + \frac{\beta D}{2\pi} \right) - \Psi\left(\frac{1}{2} + \frac{\beta D}{2\pi} + i\beta\frac{\veps_b - \veps_t -\mu_{\alpha b} + \mu_{Lt}}{2\pi}\right) \right) \right]
\right\} \,,
\end{multline}
\beq
\gamma_{tb}^{\alpha bRt} 
= \frac{\beta\Gamma_{\alpha b}\Gamma_{Rt}}{4\pi^2\hbar} n_B(\veps_b - \veps_t + \mu_{Rt}-\mu_{\alpha b})
\Im\left[\Psi^{(1)}\left(\frac{1}{2} + i\beta\frac{\veps_b-\mu_{\alpha b}}{2\pi}\right)
- \Psi^{(1)}\left(\frac{1}{2} + i\beta\frac{\veps_t-\mu_{Rt}}{2\pi}\right)\right] \,,
\edq
\beq
\gamma_{bt}^{Rt\alpha b} 
= \frac{\beta\Gamma_{Rt}\Gamma_{\alpha b}}{4\pi^2\hbar} n_B(\veps_t - \veps_b + \mu_{\alpha b}-\mu_{Rt})
\Im\left[\Psi^{(1)}\left(\frac{1}{2} + i\beta\frac{\veps_t-\mu_{Rt}}{2\pi}\right)
- \Psi^{(1)}\left(\frac{1}{2} + i\beta\frac{\veps_b-\mu_{\alpha b}}{2\pi}\right)\right] \,.
\edq
\eds

Unlike Ref.~\cite{Keller16}, we here obtain the expression
for the rate $\gamma_{00/bb}^{\alpha b\beta b}$ in order to
compute $I_{\rm drag}$.

\section{Drag and Drive Currents in the Sequential Tunneling Limit} \label{app:Currents}

In the stationary limit, Eqs.~\eqref{eq:fpopupd} in combination with the probability conservation law yield
\bes
\beq
P_0^{\text{st}} = \frac{\calw_{t0}\calw_{b0} + \gamma_{bt}\calw_{t0} + \gamma_{tb}\calw_{b0}}{\cald} \,,
\edq
\beq
P_t^{\text{st}} = \frac{\calw_{b0}\calw_{0t} + \calw_{0b}\gamma_{bt} + \gamma_{bt}\calw_{0t}}{\cald} \,,
\edq
\beq
P_b^{\text{st}} = \frac{\calw_{t0}\calw_{0b} + \calw_{0t}\gamma_{tb} + \gamma_{tb}\calw_{0b}}{\cald} \,,
\edq
\label{eq:probcot}
\eds
where $\cald=(\calw_{0b}+\calw_{0t})(\gamma_{bt}+\gamma_{tb}) + (\gamma_{bt}+\calw_{0b})\calw_{t0} + (\calw_{0t}+\calw_{t0}+\gamma_{tb})\calw_{b0}$. In Eqs.~\eqref{eq:probcot},
$\calw_{t0} = \sum_{\alpha}\calw_{t0}^{\alpha t}$,
$\calw_{b0} = \sum_{\alpha}\calw_{b0}^{\alpha b}$,
$\calw_{0t} = \sum_{\alpha}\calw_{0t}^{\alpha t}$, and
$\calw_{0b} = \sum_{\alpha}\calw_{0b}^{\alpha b}$
are the total sequential rates whilst
$\gamma_{tb} = \sum_{\alpha,\beta} \gamma_{tb}^{\alpha b\beta t}$ and
$\gamma_{bt} = \sum_{\alpha,\beta} \gamma_{bt}^{\alpha t\beta b}$
denote the total cotunneling rates.

To emphasize the role of the cotunneling processes in our three-state model, we now consider only the sequential rates.
The stationary probabilities then reduce to
\bes
\beq
P_0^{\text{seq,st}} = \frac{\calw_{t0}\calw_{b0}}{\calw_{0b}\calw_{t0} + \calw_{0t}\calw_{b0} + \calw_{t0}\calw_{b0}} \,,
\edq
\beq
P_t^{\text{seq,st}} = \frac{\calw_{b0}\calw_{0t}}{\calw_{0b}\calw_{t0} + \calw_{0t}\calw_{b0} + \calw_{t0}\calw_{b0}} \,,
\edq
\beq
P_b^{\text{seq,st}} = \frac{\calw_{t0}\calw_{0b}}{\calw_{0b}\calw_{t0} + \calw_{0t}\calw_{b0} + \calw_{t0}\calw_{b0}} \,.
\edq
\label{eq:proseq}
\eds
The drag current due to the sequential tunneling [see Eq.~\eqref{eq:FUIdrag} in the absence of the cotunneling processes] becomes
\begin{align}
I_{\text{drag}} 
&= -e\left(\calw_{0t}^{Lt}P_0^{\text{seq,st}} - \calw_{t0}^{Lt}P_t^{\text{seq,st}}\right) \nonumber \\
&= -e\frac{\left(\calw_{t0}^{Rt}\calw_{0t}^{Lt} - \calw_{t0}^{Lt}\calw_{0t}^{Rt}\right)\calw_{b0}}{\calw_{0b}\calw_{t0} + \calw_{0t}\calw_{b0} + \calw_{t0}\calw_{b0}} \,.
\end{align}
Using Eqs.~\eqref{eq:seq0i} and \eqref{eq:seqi0}, we find Eq.~\eqref{eq_Idrag}.
Following the same reasoning, it can be shown that the drive current becomes
\begin{align}
I_{\text{drive}} &\propto \Gamma_{Lb}(\veps_b)\Gamma_{Rb}(\veps_b)\nonumber \\
&\times\left\{\left[1-f_{Rb}(\veps_b)\right]f_{Lb}(\veps_b) - \left[1-f_{Lb}(\veps_b)\right]f_{Rb}(\veps_b)\right\}  \,.
\end{align}
In this case, the left and right bottom reservoirs have different chemical potentials [i.e., $f_{Lb}(\veps_b) \ne f_{Rb}(\veps_b)$], thereby the drive current does not vanish unlike the sequential drag current.
We conclude that the contunneling processes are the essential ingredient to get a drag current in the minimal three-state model.
This is in contrast to the four-state model in which even the sequential tunneling processes alone can induce a drag current \cite{Sanchez10}.

\section{Electrostatic Model} \label{app:EM}

\begin{figure}[t]
\centering
\includegraphics[width=0.45\textwidth,angle=0]{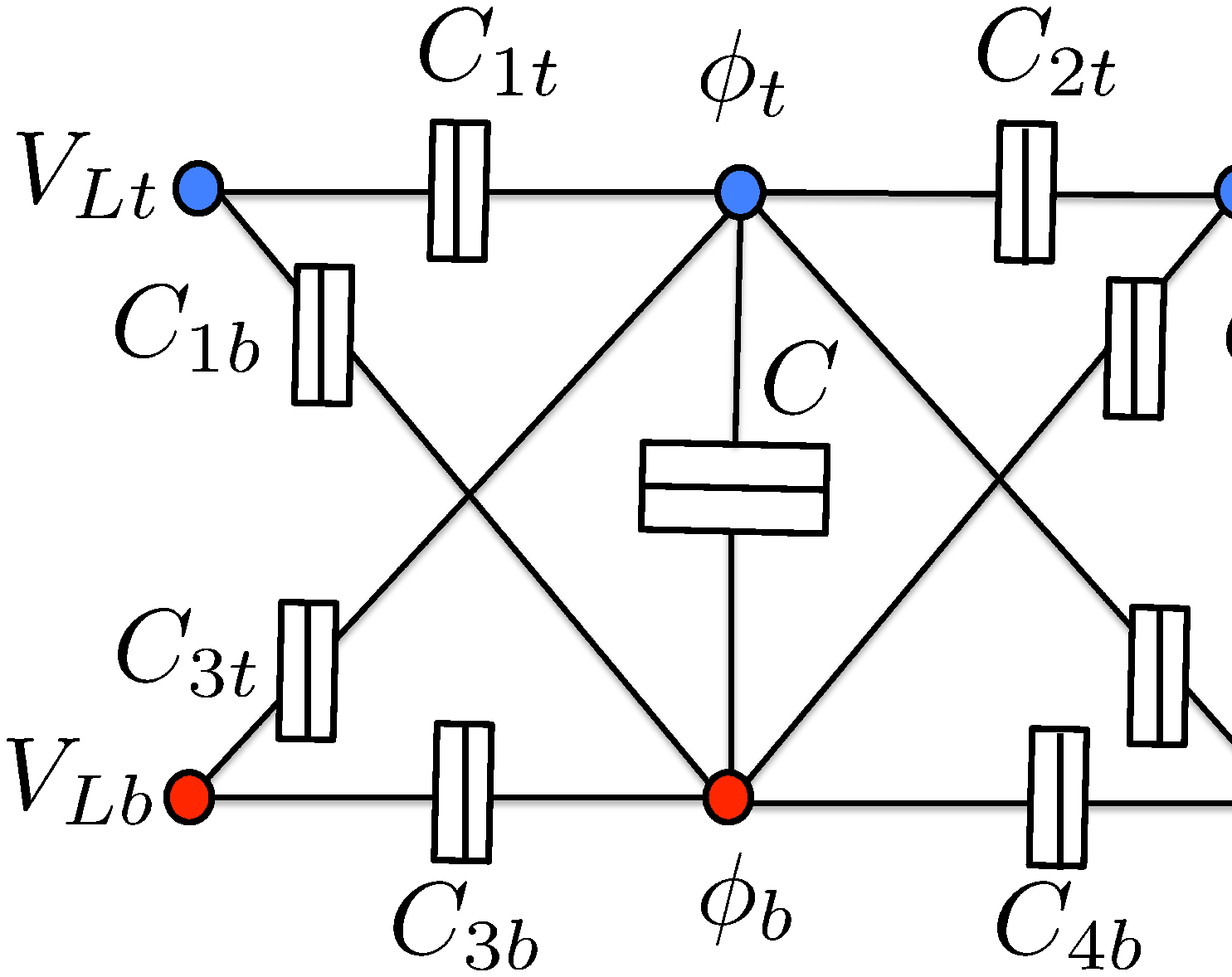}
\caption{Electrostatic model for the double quantum dot system.}
\label{fig:capacitor}
\end{figure}

With the geometry shown in Fig.~\ref{fig:capacitor}, the electrostatic equations for the charges $Q_{t}$ and $Q_{b}$ are given by
\bes\label{eq_qtqb}
\begin{align}
Q_{t} &= \sum_{\alpha=1}^4 C_{\alpha t}(\phi_{t} - V_{\alpha}) + C(\phi_t - \phi_b) \,, 
\\
Q_{b} &= \sum_{\alpha=1}^4 C_{\alpha b}(\phi_{b} - V_{\alpha}) + C(\phi_b - \phi_t)\,.
\end{align}
\eds
To shorten the notation, we set $V_1=V_{Lt}$, $V_2=V_{Rt}$, $V_3=V_{Lb}$ and $V_4=V_{Rb}$.
Solving Eqs.~\eqref{eq_qtqb}, we find the internal potentials $\phi_{t}$ and $\phi_{b}$
\bes
\begin{align}
\phi_{t} &= \frac{1}{K}\left[\sum_{\alpha} C_{\alpha b} (Q_{t} + \sum_{\beta} C_{\beta t}V_{\beta}) 
\right. \left. \right.
+ C(Q_{t} + Q_{b}  
+ \left. \sum_{\alpha}\sum_{i=t/b} C_{\alpha i}V_{\alpha})\right] \,,
\end{align}
\begin{align}
\phi_{b} &= \frac{1}{K}\left[\sum_{\alpha} C_{\alpha t} (Q_{b} + \sum_{\beta} C_{\beta b}V_{\beta}) 
\right. \left.\right.
+ C(Q_{t} + Q_{b} 
+ \left.\sum_{\alpha}\sum_{i=t/b} C_{\alpha i}V_{\alpha})\right] \,,
\end{align}
\eds
where
\beq
K = \sum_{\alpha} C_{\alpha t}\sum_{\beta} C_{\beta b} + C\sum_{\alpha} \sum_{i=t/b} C_{\alpha i} \,.
\edq
The potential energies with $N_{t}$ and $N_{b}$ excess electrons then take the form
\bes
\beq
U_{t}(N_{t},N_{b}) = \int_0^{eN_{t}} dQ_{t}~ \phi_{t}(Q_{t},Q_{b}) \,,
\edq
\beq
U_{b}(N_{t},N_{b}) = \int_0^{eN_{b}} dQ_{b}~ \phi_{b}(Q_{t},Q_{b}) \,,
\edq
\eds
which become
\bes
\begin{align}
U_{t}(N_{t},N_{b}) &= \frac{eN_t}{2K}\left[\sum_{\alpha} C_{\alpha b} (eN_t + 2\sum_{\beta} C_{\beta t}V_{\beta}) 
\right.  \left.\right. 
+ \left. C(eN_t + 2eN_b + 2\sum_{\alpha}\sum_{i=t/b} C_{\alpha i}V_{\alpha})\right] \,,
\end{align}
\begin{align}
U_{b}(N_{t},N_{b}) &= \frac{eN_b}{2K}\left[\sum_{\alpha} C_{\alpha t} (eN_b + 2\sum_{\beta} C_{\beta b}V_{\beta}) 
\right.  \left. \right.
+ \left. C(2eN_t + eN_b + 2\sum_{\alpha}\sum_{i=t/b} C_{\alpha i}V_{\alpha})\right] \,.
\end{align}
\eds

In the uncharged case, the electrochemical potential (effective level) of top or bottom dot can be written as
\bes\label{eq_ref}
\beq
\veps_{t0}^{\rm eff} = E_{t} - E_0 = \veps_{t} + U_{t}(1,0) - U_{t}(0,0) \,,
\edq
\beq
\veps_{b0}^{\rm eff} = E_{b} - E_0 = \veps_{b} + U_{b}(0,1) - U_{b}(0,0) \,,
\edq
\eds
where $E_i$ is the single-particle kinetic energy plus the internal potential energy.
Equation~\eqref{eq_ref} is the energy required to add one electron into top or bottom dot when both levels are empty:
\bes
\begin{align}
\veps_{t0}^{\rm eff} &= \veps_t + \frac{e}{2K}\left[\sum_{\alpha} C_{\alpha b} (e + 2\sum_{\beta} C_{\beta t}V_{\beta}) 
\right.  \left.\right. 
+ \left. C(e + 2\sum_{\alpha}\sum_{i=t/b} C_{\alpha i}V_{\alpha})\right] \,,
\end{align}
\begin{align}
\veps_{b0}^{\rm eff} &= \veps_b + \frac{e}{2K}\left[\sum_{\alpha} C_{\alpha t} (e + 2\sum_{\beta} C_{\beta b}V_{\beta}) 
\right. \left.\right. 
+ \left.  C(e + 2\sum_{\alpha}\sum_{i=t/b} C_{\alpha i}V_{\alpha})\right] \,.
\end{align}
\eds
The Fermi functions are evaluated at $\veps_{i0}^{\rm eff} - \mu_{\alpha}$, where $\mu_{\alpha} = E_F + eV_{\alpha}$ with $E_F$ the common Fermi energy:
\bes
\begin{align}
\veps_{t0}^{\rm eff} - \mu_1 &= \veps_t - E_F 
+ \frac{1}{K}\left[\frac{e^2\left(\sum_{\alpha}C_{\alpha b} + C\right)}{2} + e\sum_{\alpha,\beta}C_{\alpha b}C_{\beta t}V_{\beta 1} + e\sum_{\alpha,i} CC_{\alpha i}V_{\alpha 1}\right] \,,
\\
\veps_{t0}^{\rm eff} - \mu_2 &= \veps_t - E_F 
+ \frac{1}{K}\left[\frac{e^2\left(\sum_{\alpha}C_{\alpha b} + C\right)}{2} + e\sum_{\alpha,\beta}C_{\alpha b}C_{\beta t}V_{\beta 2} + e\sum_{\alpha,i} CC_{\alpha i}V_{\alpha 2}\right] \,,
\\
\veps_{b0}^{\rm eff} - \mu_3 &= \veps_b - E_F 
+ \frac{1}{K}\left[\frac{e^2\left(\sum_{\alpha}C_{\alpha t} + C\right)}{2} + e\sum_{\alpha,\beta}C_{\alpha t}C_{\beta b}V_{\beta 3} + e\sum_{\alpha,i} CC_{\alpha i}V_{\alpha 3}\right] \,,
\\
\veps_{b0}^{\rm eff} - \mu_4 &= \veps_b - E_F 
+ \frac{1}{K}\left[\frac{e^2\left(\sum_{\alpha}C_{\alpha t} + C\right)}{2} + e\sum_{\alpha,\beta}C_{\alpha t}C_{\beta b}V_{\beta 4} + e\sum_{\alpha,i} CC_{\alpha i}V_{\alpha 4}\right] \,,
\end{align}
\eds
Here, $V_{\alpha\beta} = V_{\alpha} - V_{\beta}$ and our model is gauge invariant as should be.

For $U\to\infty$, the doubly occupied state is forbidden, which amounts to taking the limit $C\to\infty$.
We then have
\bes
\begin{align}
\veps_{t0}^{\rm eff} - \mu_1 &= \veps_t - E_F 
+ \frac{e^2}{2C_{\Sigma}} + \frac{e}{C_{\Sigma}}\sum_{\alpha,i} C_{\alpha i}V_{\alpha 1}\,,
\\
\veps_{t0}^{\rm eff} - \mu_2 &= \veps_t - E_F 
+ \frac{e^2}{2C_{\Sigma}} + \frac{e}{C_{\Sigma}}\sum_{\alpha,i} C_{\alpha i}V_{\alpha 2}\,,
\\
\veps_{b0}^{\rm eff} - \mu_3 &= \veps_b - E_F 
+ \frac{e^2}{2C_{\Sigma}} + \frac{e}{C_{\Sigma}}\sum_{\alpha,i} C_{\alpha i}V_{\alpha 3}\,,
\\
\veps_{b0}^{\rm eff} - \mu_4 &= \veps_b - E_F 
+ \frac{e^2}{2C_{\Sigma}} + \frac{e}{C_{\Sigma}}\sum_{\alpha,i} C_{\alpha i}V_{\alpha 4}\,
\end{align}
\eds
with
\beq
C_{\Sigma} = \sum_{\alpha,i} C_{\alpha i} \,.
\edq
For the numerical simulations of Sec.~\ref{sec_results} we use in units of $e^2/\Gamma_0$ the values
$C_{1t}=1/202$, $C_{2t}=1/815$, $C_{3b}=1/1485$, $C_{4b}=1/177$ and
$C_{1b}=C_{2b}=C_{3t}=C_{4t}=0$. These values have been extracted from
the experimental stability diagrams of Ref.~\cite{Keller16}, as reported in its Supplementary Material~\cite{suppl}.

\end{document}